# Non-parametric attenuation curves in local star-forming galaxies: geometry effect, dust evolution, and intermediate-scale structures

Jiafeng Lu(卢家风)[1]★ Xi Kang(康熙)[1,2,3]★ Shiyin Shen(沈世银)[4,5] Qi Zeng(曾琪)[4,6,7] and Shuai Feng(冯帅)[8,9,10]

[1] *Institute for Astronomy, School of Physics, Zhejiang University, Hangzhou 310027, China*
[2] *Center for Cosmology and Computational Astrophysics, Zhejiang University, Hangzhou 310027, China*
[3] *Purple Mountain Observatory, 10 Yuan Hua Road, Nanjing 210034, China*
[4] *Shanghai Astronomical Observatory, Chinese Academy of Sciences, 80 Nandan Road, Shanghai 200030, China*
[5] *Key Lab for Astrophysics, Shanghai 200034, China*
[6] *School of Astronomy and Space Sciences, University of Chinese Academy of Sciences, No. 19A Yuquan Road, Beijing 100049, China*
[7] *Sterrenkundig Observatorium, Universiteit Gent, Krijgslaan 299, B-9000 Gent, Belgium*
[8] *College of Physics, Hebei Normal University, 20 South Erhuan Road, Shijiazhuang 050024, China*
[9] *Shijiazhuang Key Laboratory of Astronomy and Space Science / Guoshoujing Institute of Astronomy, Hebei Normal University, Shijiazhuang, Hebei 050024, China*
[10] *Hebei Key Laboratory of Photophysics Research and Application, Shijiazhuang 050024, China*

## ABSTRACT

We introduce a non-parametric approach, the Stellar Population Synthesis with Equivalent Widths (SEW) method, to reconstruct spectrally-resolved attenuation curves for 169 568 star-forming galaxies from the Sloan Digital Sky Survey Data Release 7 (SDSS DR7). Composite attenuation curves, stacked by stellar mass and inclination, reveal systematic trends: a higher stellar mass correlates with steeper slopes (lower $R_V$), while edge-on galaxies exhibit flatter curves due to geometric saturation effects. This flattening occurs because, as optical depth increases along the line of sight, the observed light becomes increasingly dominated by emission from the outer, less obscured layers of the galaxy. Using a simplified radiative transfer treatment based on a uniform dust-star mixture, we find the inclination-dependent slope variations are consistent with geometric effects, whereas the mass-dependent slope steepening indicates evolution in intrinsic dust properties, suggesting feedback-driven grain fragmentation in massive galaxies. Additionally, intermediate-scale structures are tentatively identified in the attenuation curves at approximately 4870, 6370, and 7690 Å. These results illustrate how the interplay among dust-star geometry, grain size evolution, and the galactic environment shapes attenuation curves.

**Key words:** ISM: dust, extinction – ISM: evolution – galaxies: evolution – galaxies: ISM.

## 1 INTRODUCTION

Interstellar dust, an important component of the interstellar medium (ISM), profoundly shapes the observable properties of galaxies. Dust grains form via stellar processes including supernova ejecta (A. S. Ferrarotti & H. P. Gail 2006) and asymptotic giant branch (AGB) stellar winds (E. Dwek & I. Cherchneff 2011; R. Indebetouw et al. 2014). They evolve through coagulation, shattering, destruction, and accretion in diverse galactic environments (F. Galliano, M. Galametz & A. P. Jones 2018). These processes regulate grain size distributions and chemical composition, modulating dust-radiation interactions (R. S. Asano et al. 2013; S. Salim & D. Narayanan 2020; R. McKinnon et al. 2021; N. H. Soliman, P. F. Hopkins & M. Y. Grudić 2024). Dust absorbs and scatters UV-optical radiation, redistributing energy into the infrared (IR). This imprints signatures on galaxy spectral energy distributions (SEDs) and affects the reliability of derived properties, such as stellar mass and star formation rates of galaxies (J. Kennicutt 1998; D. A. Gadotti, M. Baes & S. Falony 2010; C. C. Popescu et al. 2011; B. A. Pastrav et al. 2013).

Interstellar dust absorption is described by two related but critically distinct concepts: extinction and attenuation. Extinction quantifies light lost along a single sight line through interstellar dust. This light loss is determined by dust grain properties and column density (J. S. Mathis 1990; B. T. Draine 2003). Attenuation, conversely, encompasses radiative transfer effects arising from the distribution of stars and dust in the galaxy. Theoretical models and simulations demonstrate that varied dust column densities and complex star-dust geometries produce diverse attenuation curves, even with fixed dust properties (J. Chevallard et al. 2013; K.-I. Seon & B. T. Draine 2016; D. Narayanan et al. 2018; X. Shen et al. 2020; J. W. Trayford et al. 2020; Y.-H. Lin et al. 2021; C. Hahn et al. 2022; M. Mushtaq et al. 2023; Y. Dubois et al. 2024; F. Di Mascia et al.

★ E-mail: jefferslu@live.com (JL); kangxi@zju.edu.cn (XK)

2025; L. Sommovigo et al.2025 ). Such geometric complexity causes systematic differences between observed attenuation and intrinsic dust extinction.

Extinction curve slopes of individual stars in the Milky Way and local group galaxies, measured by the pair method and characterized by parameters like the total-to-selective attenuation $R_V$, exhibit significant variations (J. A. Cardelli, G. C. Clayton & J. S. Mathis 1989; E. L. Fitzpatrick 1999). Besides, recent Milky Way studies reveal intermediate-scale structure (ISS) in extinction curves (D. Massa, E. L. Fitzpatrick & K. D. Gordon 2020; R. Zhang et al. 2024). However, limited spatial resolution in distant extra-galaxies prevents direct extinction measurement, making attenuation the observable dust signature. Traditional approaches to derive attenuation curves fall into two categories: the comparison method and parametric SED modelling. The comparison method, analogous to the pair method, relies on empirical templates of 'unattenuated' galaxies (e.g. D. Calzetti, A. L. Kinney & T. Storchi-Bergmann 1994; D. Calzetti et al. 2000; V. Wild et al. 2011; A. J. Battisti, D. Calzetti & R. R. Chary 2016, 2017). In contrast, model-based methods employ stellar population synthesis codes to fit observed SEDs with parametrized attenuation curves, allowing per-galaxy constraints (e.g. M. Kriek & C. Conroy 2013; B. Salmon et al. 2016; J. Leja et al. 2017; S. Salim, M. Boquien & J. C. Lee 2018; I. Shivaei et al. 2020; M. Boquien et al. 2022; M. Hamed et al. 2023; B. Lorenz et al. 2023; V. Markov et al. 2023, 2025; R. Fisher et al. 2025). Different method approaches and sample selections systematically yield divergent attenuation curve slopes, ranging from Calzetti-like shallow slopes to steeper SMC-like curves.

The limitations of traditional methods arise from their inherent assumptions. The comparison method requires that the comparison sample either be dust-free or share the same attenuation curve as the target galaxy. This approach can be applicable under specific dust-star geometries (e.g. screen-like geometry), where the curve shape shows a weaker first-order dependence on total optical depth Under such conditions, galaxies with different optical depth can still share a similar curve shape, allowing them to be used as comparative samples (e.g. N. A. Reddy et al. 2015, 2018 use this method in high-redshift galaxies as their clumpy distribution). In contrast, for galaxies with a more homogeneous or mixed dust-star geometry, the attenuation curve becomes a strong function of optical depth. Consequently, the comparison method conflates geometric effects with intrinsic dust properties, as galaxies classified as 'unattenuated' often still contain substantial dust obscuration (S. Salim et al. 2018). Parametric SED modelling suffers from degeneracies among stellar ages, metallicities, and dust parameters due to assuming a global attenuation curve (C. Conroy, M. White & J. E. Gunn 2010; J. Qin et al. 2022; J. Lu et al.2025). Moreover, geometry effect produce non-linear relations between observed attenuation and intrinsic dust properties, challenging parametric models (S. Charlot & S. M. Fall 2000; J. Chevallard et al. 2013). Observed ISSs in the Milky Way extinction curve [i.e. localized features in wavelength space between diffuse interstellar bands (DIBs) and broad continuum variations] further challenge parametric attenuation prescriptions, potentially biasing derived slopes (D. Massa et al. 2020; R. Zhang et al. 2024).

These limitations underscore the need for non-parametric approaches to reconstruct attenuation curves without relying on prior assumptions regarding their shape or on unattenuated spectral templates. Absorption line equivalent width, insensitive to flux calibration bias and attenuation, can be used to infer the star formation history (SFH), thereby providing intrinsic unattenuated spectra. However, the non-linear nature of the equation involving equivalent width, which is ratio quantities, makes it difficult to accurately recover the SFH. To overcome this, I. Barišić et al. (2020) employ pre-defined templates for compare, while Y.-H. Lin et al. (2021) use principal component analysis to reduce the dimensionality of the spectral features, both aiming to approximate the SFH.

Emerging techniques, such as the Stellar population synthesis with Equivalent Widths (SEW) method (J. Lu et al. 2025), overcome this challenge by leveraging equivalent widths, which are insensitive to dust attenuation, to simultaneously and self-consistently derive both the intrinsic stellar spectrum and the non-parametric attenuation curve. This method yields non-parametric attenuation curves, enabling spectral-resolution analysis and radiative transfer modelling from extinction to attenuation.

In this work, we apply the SEW method to derive non-parametric attenuation curves for a sample of star-forming galaxies (SFGs) from the Sloan Digital Sky Survey (SDSS). We then investigate the dependence of attenuation slopes on both stellar mass and galactic inclination, recognizing stellar mass as a critical scaling parameter for galaxy evolution and inclination as a key factor affecting dust attenuation in disc galaxies. Finally, using radiative transfer with a uniform dust-star mixture model, we examine the mass-dependent evolutionary trends of dust properties and their underlying physical mechanisms of dust evolution.

The paper is organized as follows: Section 2 describes sample selection, SEW fitting and attenuation curve stacking. Section 3.1 presents the evolutionary trends of attenuation curves with stellar mass and inclination. Section 3.2 interprets trends via radiative transfer with uniform mixed geometry. Section 4 discusses the ISSs and implications for dust evolution. Finally, we summarize our conclusions in Section 5.

## 2 DATA AND METHODOLOGY

### 2.1 SDSS sample selection

In this study, we investigate the central dust attenuation curves of disc-dominated SFGs in the local universe. Our fiber spectrum sample is selected from the Main Galaxy Sample (MGS) of the Sloan Digital Sky Survey Data Release 7 (SDSS DR7).[1] Stellar mass measurements and emission line fluxes (H $\alpha$, H $\beta$, [O III]$\lambda$5007, and [N II]$\lambda$6584) are obtained from the MPA-JHU data release[2] (G. Kauffmann et al. 2003a; J. Brinchmann et al. 2004; C. A. Tremonti et al. 2004), while galaxy inclination angles ($\theta$) are derived from L. Simard et al. (2011).

We initially identify SFGs through their emission line characteristics, employing the classical Baldwin–Phillips–Terlevich diagram (J. A. Baldwin, M. M. Phillips & R. Terlevich 1981) to exclude galaxies exhibiting active galactic nucleus signatures. Specifically, we apply the criteria outlined in G. Kauffmann et al. (2003b), we require signal-to-noise ratios (S/N) greater than 3 for all four emission lines. To ensure that the sample is dominated by disc galaxies, whose inclination angles $\theta$ satisfy a uniform distribution in $\cos\theta$ (S. P. Driver et al. 2007), we restrict the sample to galaxies with stellar mass between $10^{9.5}$ and $10^{11}$ $M_{\odot}$. As shown in the bottom left of Fig. 1, the sample galaxies have a roughly uniform distribution

[1] http://skyserver.sdss.org/dr7/

[2] https://wwwmpa.mpa-garching.mpg.de/SDSS/DR7/

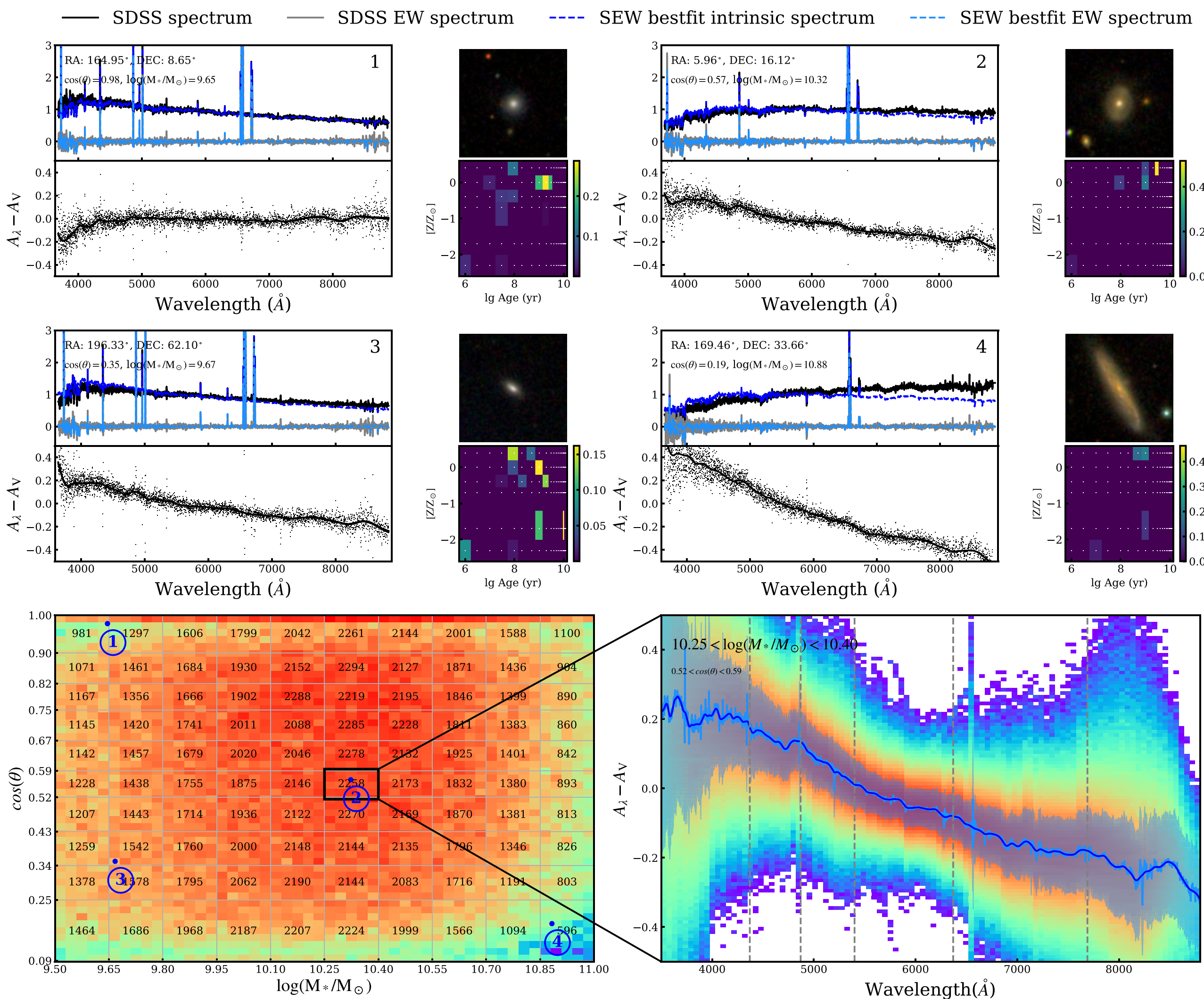

**Figure 1.** SEW fitting, binning methodology and composite attenuation curve construction. Top two rows of panels: Illustration of SEW fitting to SDSS spectra. In each subplot, the top right shows the SDSS image of the selected galaxy; the top left displays the observed spectrum, the fitted intrinsic spectrum, and their corresponding equivalent width spectra; the bottom left presents the non-parametric attenuation curve derived from the fit; and the bottom right shows the resulting stellar population distribution. Bottom left panel: Density distribution of galaxies in stellar mass-inclination space with grey binning grid. Bottom right panel: An example bin in mass range $10.25 < \log(M_*/\mathrm{M}_\odot) < 10.4$ and inclination range $0.52 < \cos\theta < 0.59$ shows the attenuation curve probability density distributions of all subsample galaxies (rainbow background), the median attenuation curve (light blue line) and its smoothed counterpart (solid blue curve) derived through DPLS. The blue shaded region represents the quartile range. Vertical grey dashed lines mark rest-frame wavelengths of key ISSs in the Milky Way: 4370, 4870, 5400, 6370, and 7690 Å.

of $\cos\theta$ at give stellar mass. The final sample comprises 169 568 SFGs.

## 2.2 SED fitting with SEW

We employ the Stellar population synthesis with Equivalent width spectrum (SEW) method to derive non-parametric attenuation curves. Developed by J. Lu et al. (2025) with inspiration from N. Li et al. (2020), the firefly (D. M. Wilkinson et al. 2015, 2017) and Lick/IDS indices (D. Burstein et al. 1984; G. Worthey et al. 1994), the SEW method employs equivalent width (obtained by the discrete penalized least squares (DPLS) method), quantities intrinsically robust to dust extinction and flux calibration uncertainties, for stellar population synthesis without a prior attenuation curve shape. The 'non-parametric' nature specifically refers to the reconstruction of attenuation curves without assuming any functional form, while stellar population weights are determined as free parameters through linear least-squares regression on the equivalent width matrix equation (equation 19 in J. Lu et al. 2025). This remains effective even in low signal-to-noise and biased calibration spectra regimes where traditional methods fail. This approach circumvents systematic biases inherent in parametric curve assumptions, while enabling concurrent analysis of three key elements: intermediate-scale attenuation features, stellar population modelling systematics, and spectral flux biases. Rigorous testing on mock spectra (J. Lu et al. 2025) has demonstrated the robustness of this method, showing accurate recovery of stellar population parameters and attenuation curves even under significant systematic biases.

Our implementation uses the PPXF-SEW PYTHON package,[3] a SEW extension package for the PPXF (M. Cappellari & E. Emsellem 2004; M. Cappellari 2017). As installations in the top two rows of Fig. 1 shows, the workflow comprises as: (1) stellar templates are derived from parametrized GALAXEV models (G. Bruzual & S. Charlot 2003) with a Chabrier IMF (G. Chabrier 2003), comprising 120 populations across 20 ages (1–13 Gyr) and six metallicities ($[M/\mathrm{M}_\odot] = -2.3$–0.4), covering 3500–9000 Å; (2) extract the equivalent with spectrum (grey lines) from the observed spectrum (black lines); (3) linear least-squares fitting of the equivalent width spectrum (equation 19 in J. Lu et al. 2025; light blue lines) to determine free parameters for stellar population weights (bottom right of each subfigure, encoding age/metallicity) and kinematics; (4) reconstruction of the intrinsic spectrum $F_{\mathrm{int}}(\lambda)$ (blue lines) from the fitted weights; and (5) direct computation of the non-parametric attenuation curve (bottom left of each sub figure):

$$A(\lambda) - A(\lambda_0) = -2.5 \log \frac{F_{\mathrm{obs}}(\lambda)/F_{\mathrm{obs}}(\lambda_0)}{F_{\mathrm{int}}(\lambda)/F_{\mathrm{int}}(\lambda_0)}, \quad (1)$$

where $\lambda_0$ is the normalization wavelength (we adopt the $V$ band at ∼5500 Å). This delivers attenuation curves as non-parametric outputs, free from functional prescriptions.

The SEW method determines relative attenuation differences by comparing the observed spectrum with the intrinsic spectrum derived from the fitted stellar weights independently of attenuation curve assumptions. This formulation resolves the degeneracy between intrinsic SEDs and attenuations, a longstanding challenge since unattenuated SEDs are unobservable. While conventional methods rely on parametric curves to constrain (introducing shape-dependent biases and age-metallicity-attenuation degeneracy; J. Qin et al. 2022; J. Lu et al. 2025), the SEW method self-consistently derives both stellar populations and dust properties without prior attenuation models. This allows us to directly investigate how attenuation curve shapes correlate with physical galaxy properties such as stellar mass and inclination, without the biases introduced by parametric assumptions.

### 2.3 Attenuation curve median stacking

The stellar mass, often used as a key scaling parameter of galaxy evolution [e.g. $M_* - Z$ (C. A. Tremonti et al. 2004), $M_* - SFR$ (J. Brinchmann et al. 2004), and Tully–Fisher relation (R. B. Tully & J. R. Fisher 1977)], reflecting the long-term evolutionary properties of galaxies. Meanwhile, the inclination of galactic disc significantly alters geometric projections and dust attenuation, and the lack of inclination corrections can lead to systematic biases. To systematically analyse dust attenuation trends, we construct composite curves across parameter space by stacking galaxies in the two key parameters.

Our analysis begins with dividing the galaxy sample into 10 equal-width stellar mass bins spanning $9.5 < \log(M_*/\mathrm{M}_\odot) < 11$ and 10 equal-sized inclination bins based on axis ratio. This orthogonal binning strategy creates a $10 \times 10$ grid containing 100 distinct subsamples. The bottom left panel illustrates the binning grid with grey lines of Fig. 1. This strategy ensures balanced sampling across both parameter dimensions. For each grid cell, we compute characteristic attenuation curves through per-wavelength median stacking of individual galaxy measurements. Uncertainties of the median stacking attenuation curves are estimated via $\sqrt{\frac{\pi}{2}} \frac{\sigma}{\sqrt{n}}$, where $\sigma$ is the standard deviation of attenuation curves distribution at a given wavelength and $n$ is the number of galaxies per bin. According to our criterion, with about at least 900 galaxies per bin (recorded in the grid of Fig. 1), the stacking reduces errors by a factor of 30.

For example, the bottom right panel of Fig. 1 demonstrates this process for galaxies with stellar mass between 10.25 and 10.4 and inclination ranging from 54° to 59° ($0.52 < \cos\theta < 0.59$). Its background shows attenuation distributions of all galaxies, with blue shading indicating the 25th–75th percentile range. The light blue line traces the median differential attenuation as a function of wavelength, while the solid blue curve represents its smoothed version using DPLS method.

Fig. 2 quantifies median attenuation differences across stellar mass and inclination, revealing systematic trends whose physical origins will be analysed in Sections 3.1 and 3.2. Overall, within the 4200–8000 Å range, these curves consistently exhibit two key characteristics: a global negative slope and distinct bump-like features identified as ISSs, as discussed in Section 4.1.

However, significant deviations appear in the UV (<4200 Å) and NIR (>8000 Å) bands. UV deviations are stronger in low-mass galaxies, while NIR bumps occur only with GALAXEV templates (Appendix B). This UV biases may reflect the knee structure (where attenuation curves are known to change slopes) (J. A. Cardelli et al. 1989; B. B. Teklu et al. 2020), or stem from incomplete stellar populations: missing young stellar populations (GALAXEV with the younger ones shows weaker deviations) or calibration bias. But, as discussed in J. Lu et al. (2025), such UV deviations minimally affect the optical bands as the optical flux is dominated by small to intermediate mass stars. NIR issues may arise from insufficient post-main-sequence stars (C. Maraston 2005, e.g. TP-AGB), where EMILES templates are more complete. We present EMILES results in Appendix B, but use GALAXEV in the main text for its better coverage of young populations. Both biased regions are excluded from quantitative analysis. In the future, more complete stellar population templates (e.g. complete young OB-type star templates, binary star templates, etc.) are expected to solve this problem, but in this work, both biased regions are excluded from the quantitative analysis.

## 3 RESULTS

### 3.1 Evolutionary attenuation curve

To systematically investigate the stellar mass- and inclination-dependent attenuation properties, we refer to the formalism of S. Charlot & S. M. Fall (2000) to adopt a power-law parametrization:

$$A(\lambda) = A_V \left(\frac{\lambda}{5500\,\text{Å}}\right)^{-\beta} = E(B-V) \frac{\left(\frac{\lambda}{5500\,\text{Å}}\right)^{-\beta}}{0.8^{-\beta} - 1}, \quad (2)$$

where the equivalent wavelength of $B$ and $V$ band is 4400 and 5500 Å, and 0.8 corresponds to their wavelength ratio (4400 Å/5500 Å). $\beta$ determines the optical attenuation slope, related to the total-to-selective attenuation ratio $R_V$ as:

$$R_V = \frac{A_V}{E(B-V)} = \frac{1}{0.8^{-\beta} - 1}. \quad (3)$$

In addition, both extinction and attenuation effects are discussed in this paper, and to avoid confusion, the subscript 'att' denotes attenuation, while 'ext' denotes extinction.

We perform Markov chain Monte Carlo (MCMC) sampling on stacked SEW-derived attenuation curves, with 150 Å windows masked around 4870, 6370, and 7690 Å to minimize contamination from potential ISSs. As demonstrated in Fig. 3 for the reference curve from Fig. 1, we obtain well-constrained posterior distributions (bottom-left panel) for $E(B-V)$ and $\beta$ of the power-law attenuation curve. The top panel compares observational data (light blue) and best-fitting models (dashed grey), with dark blue segments mark spectral regions incorporated in the fitting. All best-fitting curves across stellar mass and inclination bins are superimposed on Fig. 2 as dashed lines in corresponding colours.

[3] https://pypi.org/project/ppxf-sew/

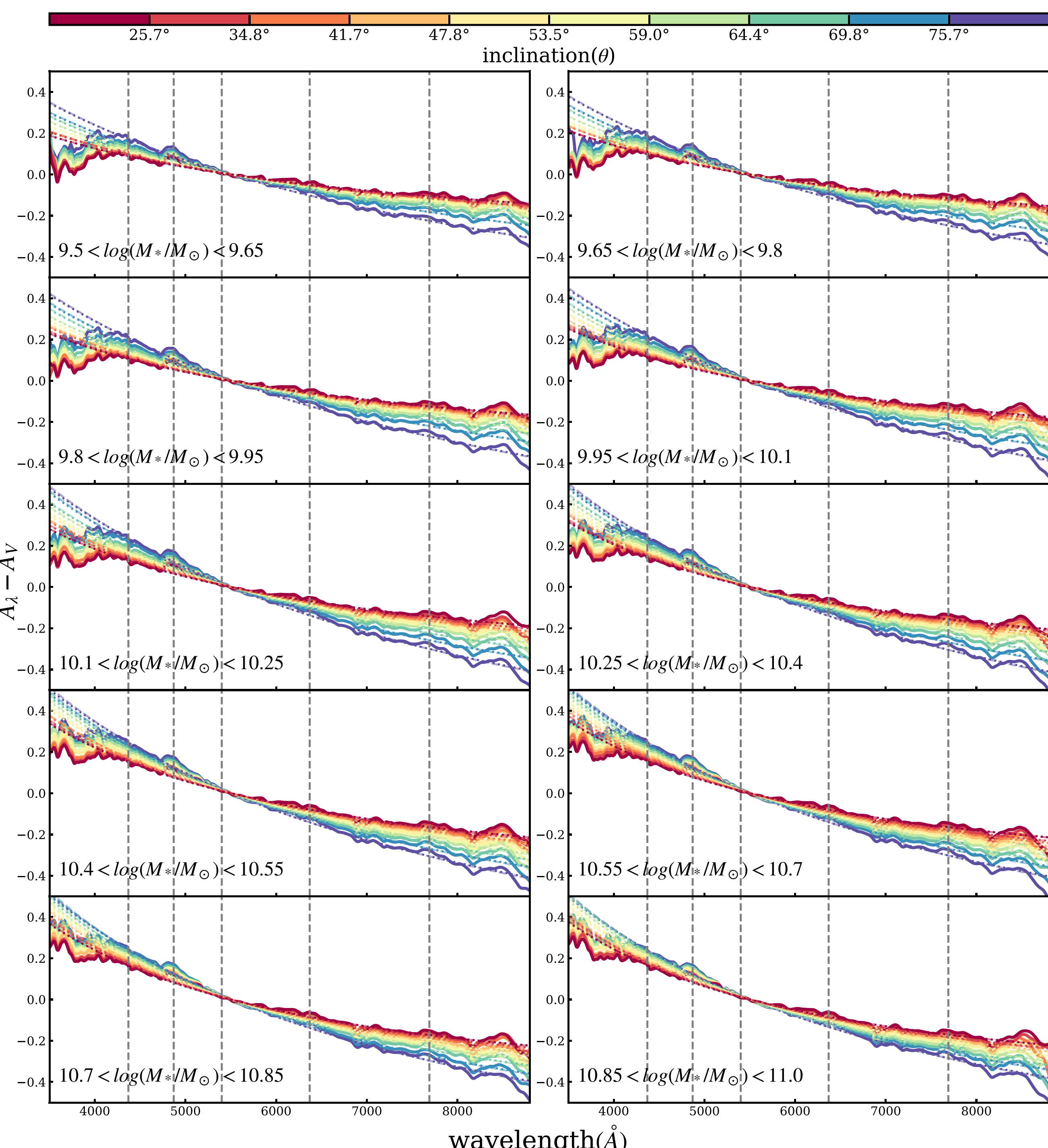

**Figure 2.** Wavelength-dependent median attenuation differences as functions of stellar mass and inclination. Each panel displays median attenuation differences for galaxies within specific stellar mass bins (annotated in the lower left corner), with inclination angles colour-coded according to the colourbar. Solid curves represent smoothed median profiles derived through SEW analysis, while dotted lines indicate the pure power-law reddening law best-fits from Section 3.1. Dashed lines correspond to the best-fitting attenuation curves combining homogeneous dust-star mixture geometry with a power-law extinction curve from Section 3.2. Vertical dotted lines mark rest-frame wavelengths of key ISSs in the Milky Way: 4370, 4870, 5400, 6370, and 7690 Å.

Fig. 4 reveals parameter correlations: upper panel displays $E_{att}(B-V)$ evolution with inclination angle, colour-coded by mass bin; lower panel shows $\beta$ versus stellar mass, colour-coded by inclination. The left $y$-axis in lower panels corresponds to $\beta_{\rm att}$, with the right $y$-axis representing its associated $R_{{\rm att},V}$ distributed non-uniformly.

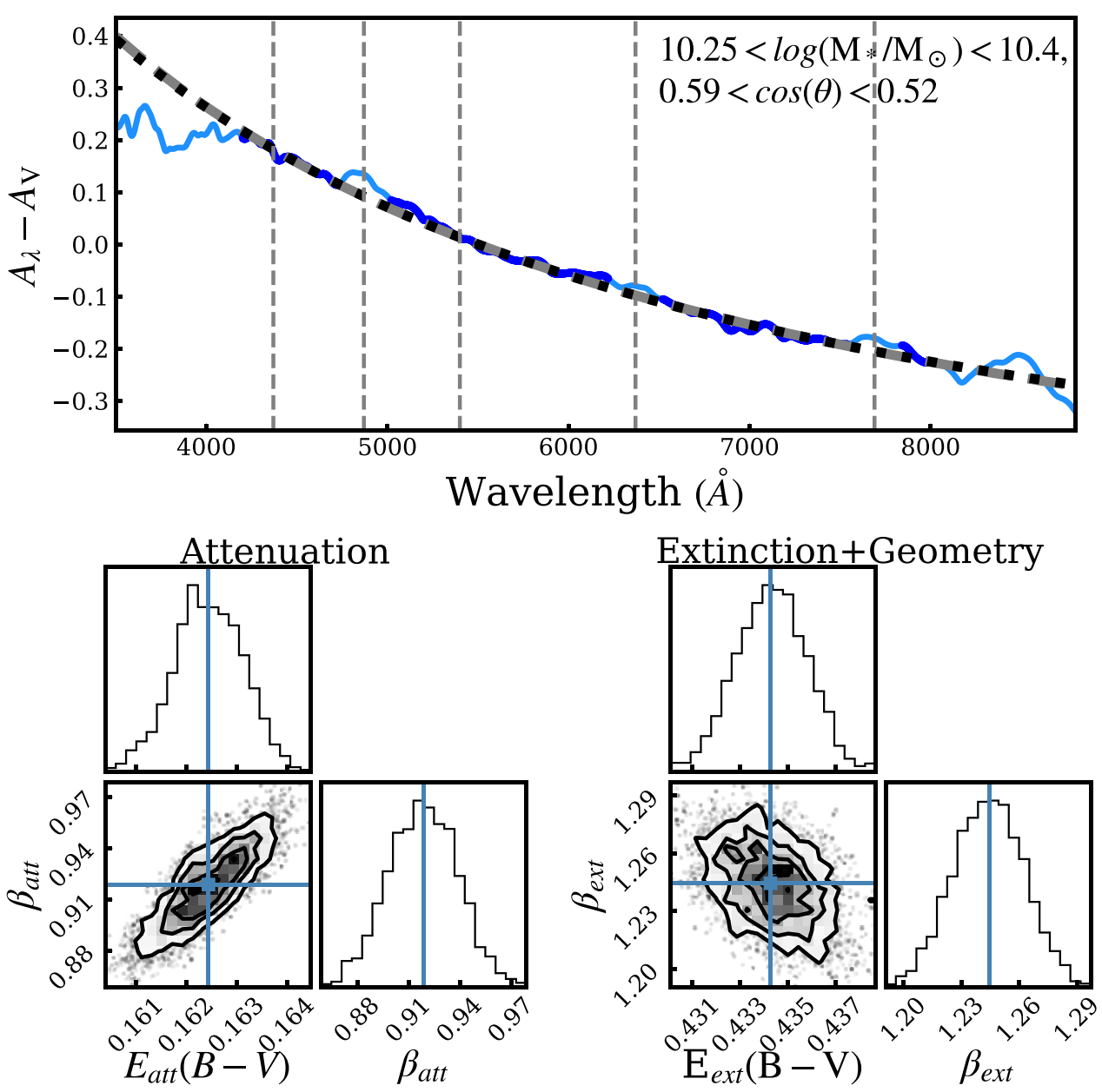

**Figure 3.** Attenuation curve fit and extinction curve with uniform mixed geometry fit for the example in the bottom panel of Fig. 1. The top panel compares the observed attenuation curve (light blue) with the best-fitting model (dashed grey for power-law attenuation curve and dotted black for extinction curve with uniform mixed geometry), where dark blue segments mark spectral regions incorporated in the fitting. The bottom-left panel shows the MCMC posterior distributions of the power-law attenuation curve parameters. While, the bottom-right panel presents the MCMC posterior distributions of extinction curve with uniform mixed geometry.

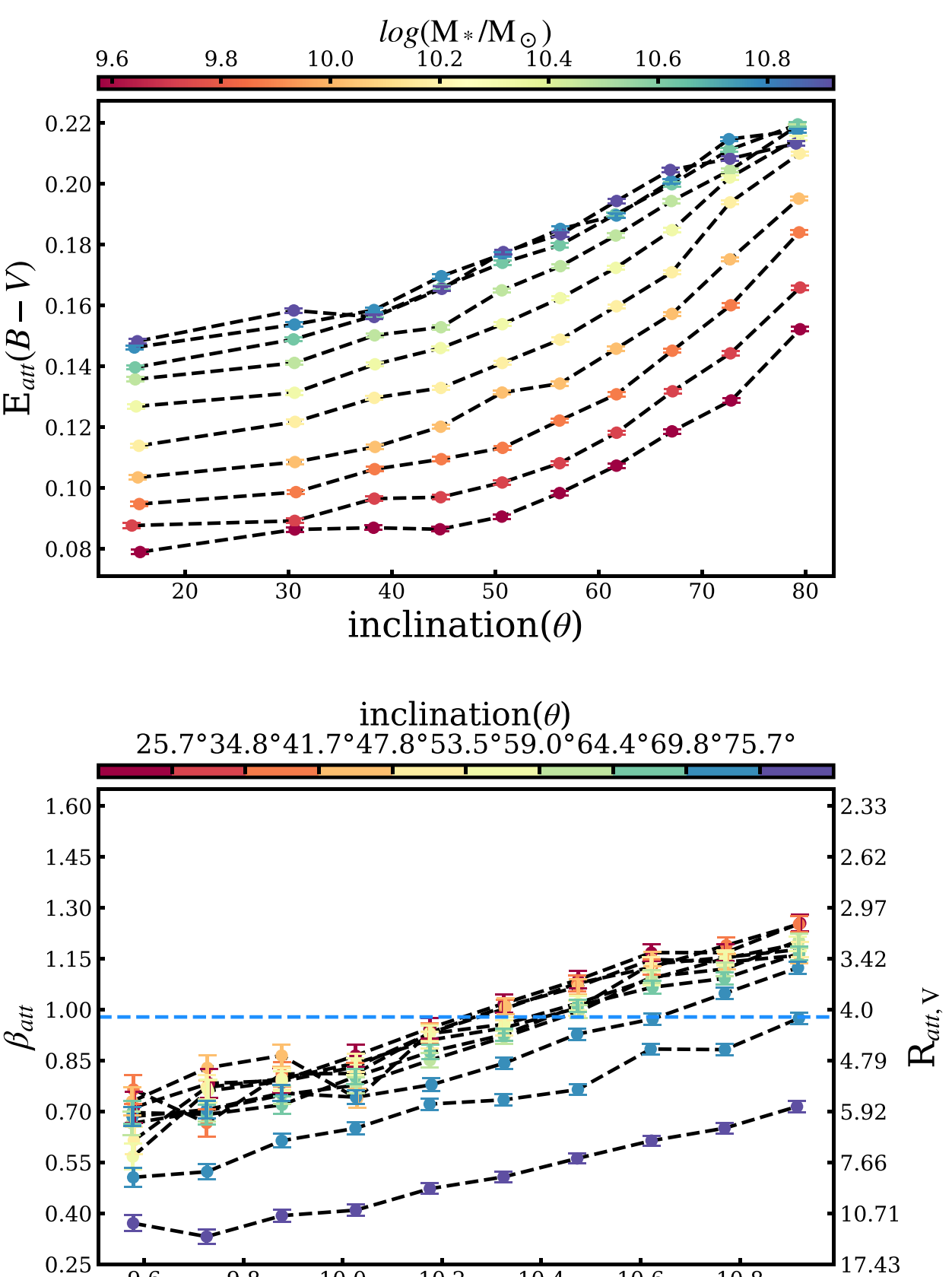

**Figure 4.** Inclination and stellar mass dependence of the best-fitting power-law attenuation curve parameters $E_{\rm att}(B-V)$ and $\beta$. Upper panel: $E_{\rm att}(B-V)$ variations with inclination angle, colour-coded according to stellar mass (greyscale bar). Lower panel: $\beta$ parameter as a function of stellar mass with inclination-colour coding, featuring dual $y$-axes showing $\beta_{\rm att}$ (left) and their corresponding $R_{{\rm att},V}$ equivalents (right). Error bars on individual points represent $1\sigma$ uncertainties derived from posterior distribution dispersions in MCMC sampling.

The $E_{\rm att}(B-V)$ parameter increases systematically with galaxy inclination. Low-mass galaxies rise from 0.07 to 0.15, while massive galaxies show an increase from 0.14 to 0.21, reaching a saturation effect at edge-on galaxies. Conversely, the attenuation slope $\beta_{\rm att}$ decreases (inverse to $R_{{\rm att},V}$ increase) with higher inclination, indicating flatter attenuation curves in edge-on galaxies. The $R_{{\rm att},V}$ of the canonical Calzetti law (D. Calzetti et al. 2000) matches it for the intermediate-mass face-on galaxies to massive intermediate inclination galaxies in our results, as shown by the blue line in Fig. 2.

The covariation of $E_{\rm att}(B-V)$ and $\beta_{\rm att}$ identify a characteristic saturation effect in galaxies where stars and dust are spatially comixed: beyond critical dust column densities, both $E_{\rm att}(B-V)$ enhancements and the flattening of attenuation profiles fundamentally trace the geometric coupling between star and dust distributions. These results align with findings by S. Salim et al. (2018) who demonstrated that galaxies with higher dust column densities systematically exhibit flattened attenuation curves, confirming the earlier conclusions of J. Leja et al. (2017) and matching predictions from radiative transfer models employing realistic three-dimensional geometries (J. Chevallard et al. 2013). Motivated by the dust-star geometry, we quantitatively characterize these geometry effects through radiative transfer models in Section 3.2.

Although $E_{\rm att}(B-V)$ correlates with both the inclination and stellar mass, $\beta$ steepens systematically with stellar mass at a fixed viewing orientation. This mass-dependent slope evolution persists across inclinations, suggesting two possibilities: either intrinsic dust property variations or differential optical depth effects between different stellar mass. We disentangle these mechanisms in Section 3.2 through uniform dust-star geometry modelling and radiative transfer, with implications for dust evolution discussed in Sections 4.3.

### 3.2 From extinction to attenuation

#### *3.2.1 Uniform dust-star mixture geometry*

Dust attenuation results from the combined effects of dust extinction and the geometric distribution of stars and dust. The latter primarily manifests in two aspects. First, it arises from spatial variations in the distribution of different radiative sources. For instance, the UV-optical attenuation slope variation observed among galaxies with different stellar mass and specific star formation rate (sSFR) can be largely attributed to this effect (N. A. Reddy et al. 2015; B. Salmon et al. 2016; S. Salim et al. 2018; J. Álvarez-Márquez et al. 2019). Young and massive stars, which dominate the UV continuum and nebular line emission, are typically embedded in regions abundant in cold gas and dust. Therefore, their emission is subject to preferentially stronger dust attenuation. As a result, galaxies with high sSFR and low mass exhibit steeper attenuation curves in the UV-optical attenuation slope, since a larger fraction of their UV emission originates from these highly attenuated clumpy young OB-type stars (X. Shen et al. 2020). Second, the geometry influences radiative transfer processes, such as stellar emission, dust absorption, and scattering, through its effect on the spatial arrangement of dust and emission sources.

In our chosen wavelength range, the clumpy attenuation effects described above, which cause differential attenuation between young and old stellar populations, do not dominate. In low-mass galaxies, young and old stars exhibit nearly identical attenuation properties, as evidenced by the consistency between Balmer decrement and stellar continuum attenuation ($E_{\rm star}(B-V)/E_{\rm gas}(B-V)\sim 1$; V. Wild et al. 2011; Y. Koyama et al. 2015; H. J. Zahid et al. 2017). In massive galaxies, however, the radiation from young stars in dusty clumps is heavily obscured, meaning that the observed attenuation primarily reflects dust in the diffuse ISM affecting the older stellar populations. On galactic scales, the dust attenuation affecting stellar emission from the diffuse ISM can be simplified and parametrized using a uniform dust-star mixture geometry model (J. Lu et al. 2022, 2023). This geometry captures the first-order effects of dust-star arrangement, though it does not include detailed scattering or clumpy structures. According to radiative transfer, the total emitted intensity $I$ is related to the total unattenuated intensity $I_0$ through:

$$I_\lambda = I_{0,\lambda}\frac{1-e^{-\tau_\lambda}}{\tau_\lambda}, \tag{4}$$

where $\tau_\lambda$ denotes the integrated optical depth along the line of sight. For a power-law dust extinction curve (equation 2), the $E_{\rm ext}(B-V)$ becomes a direct tracer of total dust column density, while the $\beta$ index reveals fundamental dust properties. The effective attenuation follows:

$$A_{{\rm att},\lambda} = -2.5\log\frac{I_\lambda}{I_{0,\lambda}} = -2.5\log\frac{1-e^{-\tau_\lambda}}{\tau_\lambda}, \tag{5}$$

where the optical depth $\tau_\lambda$ is linked to the extinction $A_{{\rm att},\lambda}$ via the relation $\tau_\lambda \equiv A_{{\rm att},\lambda}/1.086$, consistent with the definitions of attenuation and optical depth in radiative transfer.

### 3.2.2 *Dust extinction properties*

We conduct identical MCMC fitting from Section 3.1 on stacked attenuation curves to determine optimal extinction parameters for the uniform mixture model. To facilitate a comparison with the power-law attenuation curves presented in Section 3.1, Fig. 3 also presents the best uniform mixture model fits for the example shown in Fig. 1. The top panel displays the best-fitting attenuation curve (geometry effects + extinction) with a black dotted line, while the bottom-right panel shows corresponding parameters of extinction curve distributions. The same fits for all mass-inclination-binned subsamples appear as colour-matched dotted curves in Fig. 2. Although the attenuation curve in this section exhibits nearly identical shapes to those in Section 3.1, their underlying physical interpretations differ fundamentally. The attenuation curve derived here represents effective attenuation resulting from the radiative transfer of intrinsic extinction curves through a uniform mixture geometry. In contrast, the curves in Section 3.1 correspond to a direct measurement of attenuation.

Following the format of Fig. 4, Fig. 5 illustrates parameter of extinction curve dependencies on stellar mass and inclination. The upper panel tracks $E_{\rm ext}(B-V)$ variations versus inclination across stellar mass bins, while the lower panel shows $\beta_{\rm ext}$ evolution with stellar mass at different inclinations. The secondary (non-linear) $y$-axis in the lower panel also converts $\beta_{\rm ext}$ to $R_{{\rm ext},V}$ using equation (3).

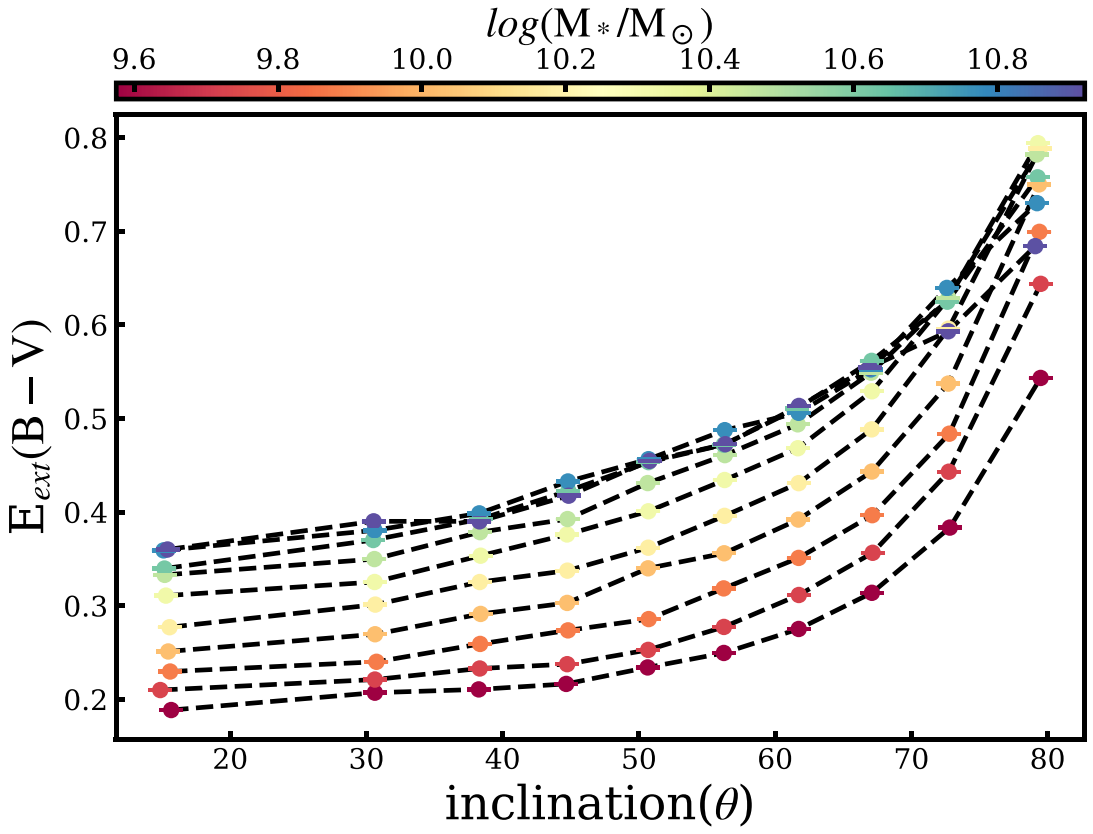

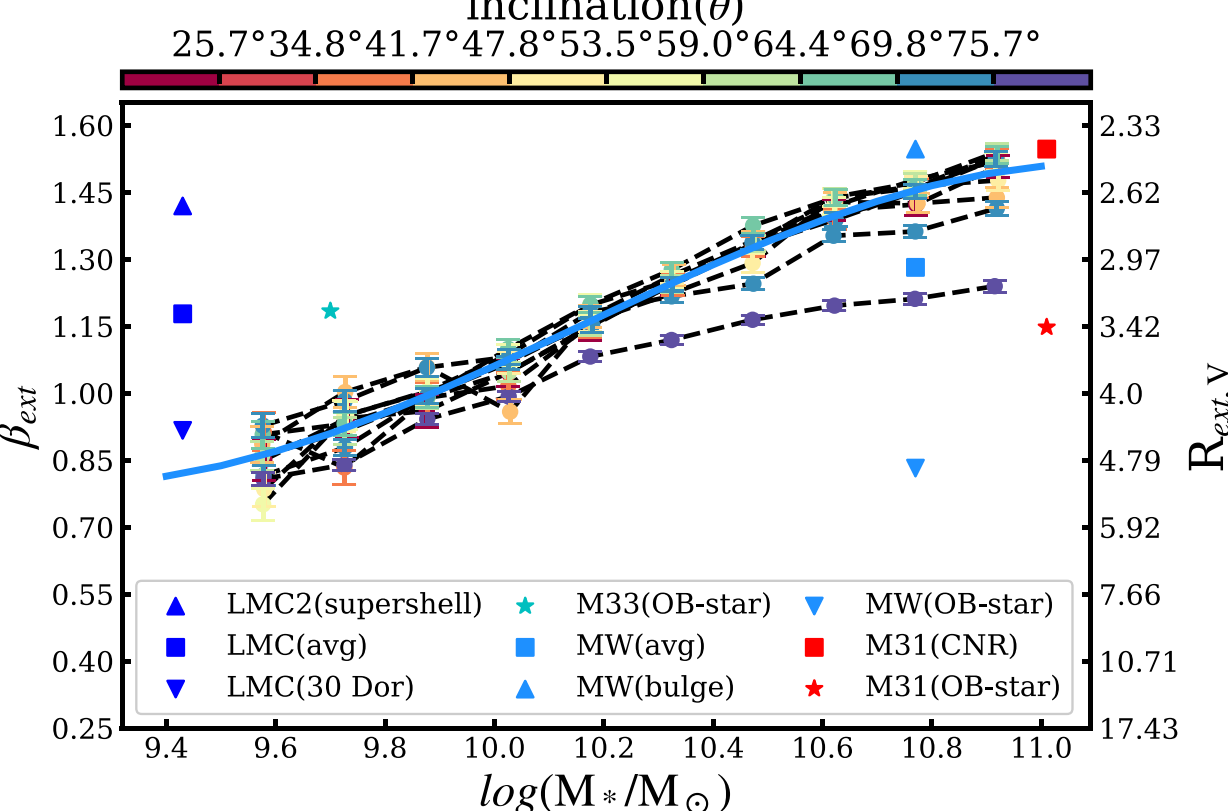

**Figure 5.** The same as Fig. 4 but for power-law extinction curve with uniform mixed geometry. In the lower panel, the extinction curve slopes of local group galaxies in different environments [The Milky Way ($\log(M_*/{\rm M}_\odot)\sim 10^{10.7}$; T. C. Licquia & J. A. Newman 2015); LMC ($\log(M_*/{\rm M}_\odot)\sim 10^{9.4}$; R. P. der Marel et al. 2002); M33 ($\log(M_*/{\rm M}_\odot)\sim 10^{9.7}$ E. Corbelli 2003; P. R. Hague & M. I. Wilkinson 2015) and M31 ($\log(M_*/{\rm M}_\odot)\sim 10^{11}$; A. Tamm et al. 2012)] are marked as legend shows.

Fig. 5 reveals crucial insights into dust geometry and extinction properties. The upper panel shows increasing $E_{\rm ext}(B-V)$ with inclination, directly reflecting sight line dust column density. This geometric dependence confirms a disc-like dust distribution where edge-on orientations lead to higher dust column density. Low-mass galaxies exhibit a significant increase in $E_{\rm ext}(B-V)$ from 0.18 (face-on) to 0.54 (edge-on), a factor of $\sim$3. Conversely, the increase is milder in massive galaxies, from 0.35 to 0.7, roughly doubling. Given that the SFGs in our sample are predominantly disc-dominated, this orientation-dependent variation primarily arises from the ratio between the vertical scale height and radial scale length of their ISM discs. This differential scaling with inclination is consistent with the presence of thickened discs in massive galaxies. The weaker inclination dependence suggests their dust distribution is more spherical, potentially resulting from bulge-driven secular evolution (J. A. Sellwood 2014, e.g. disc heating) that vertically puffs up the diffuse ISM dust distribution.

The lower panel reveals a fundamental result: the dust extinction slope exhibits a systematic steepening with stellar mass, yet remains remarkably invariant with inclination (for $\theta < 75^\circ$). This inclination independence is a key prediction of the uniform dust-star mixture model and indicates that the observed variations in the attenuation curve are primarily geometric in origin, rather than due to changes in the intrinsic dust properties. The physical origin and implications of this inclination invariance, as well as the nature of the deviations at high inclinations, are discussed in detail in Section 4.2. The mass-dependent trend (for $\theta < 75^{\circ}$) is represented by the blue curve in Fig. 5 and is described by the polynomial parametrizations (with $x = \log M_*/\mathrm{M}_{\odot} - 10$):

$$\beta_{\rm ext} = -0.22x^3 + 0.11x^2 + 0.56x + 1.06 \quad \text{or,}$$
$$R_{55,\rm ext} = 0.24x^3 + 0.67x^2 - 2.14x + 3.75. \tag{6}$$

The observed trends, decreasing $R_{\mathrm{ext},V}$ with stellar mass but invariance with inclination, also persist when using CCM extinction curve (Cardelli–Clayton–Mathis extinction curve; J. A. Cardelli et al. 1989; Appendix A; Fig. A1), despite minor parametric differences. In Section 4.3, we further explore the physical mechanisms driving this mass dependence, linking it to evolution in the dust grain size distribution.

### 3.2.3 *Saturation effect*

In the uniform mixture model, the dust colour excess $E_{\rm att}(\lambda_1 - \lambda_2)$ (i.e. reddening of attenuation) then follows:

$$E_{\rm att}(\lambda_1 - \lambda_2) = 2.5\log\left(\frac{\tau_{\lambda_1}}{\tau_{\lambda_2}}\frac{1-e^{-\tau_{\lambda_2}}}{1-e^{-\tau_{\lambda_1}}}\right), \tag{7}$$

and the $R_{\mathrm{att},V}$ is:

$$R_{\mathrm{att},V} = \frac{A_{\mathrm{att},V}}{E_{\rm att}(B-V)}. \tag{8}$$

As illustrated by equations (4) and (7), the uniform mixture model reveals that under optically thick conditions ($\tau \gg 1$), the emitted intensity follows an inverse correlation with optical depth ($I \approx I_0/\tau$), corresponding to a logarithmic dependence on the effective attenuation:

$$A_{\mathrm{att},V} \approx 2.5\log\tau_V. \tag{9}$$

The colour excess $E_{\rm att}(\lambda_1 - \lambda_2)$ becomes approximately constant, expressed as $2.5\log(\tau_{\lambda_1}/\tau_{\lambda_2})$. For extinction curves following a power law, the maximum reddening satisfies:

$$E_{\rm att}(\lambda_1 - \lambda_2) \approx 2.5\beta_{\rm ext}\log\lambda_2/\lambda_1. \tag{10}$$

This theoretical upper limit explains the observed plateau in reddening $E_{\rm att}(B-V)$ for massive edge-on galaxies shown in Fig. 4, which we term dust attenuation saturation. The non-linear correlation between $A_{\mathrm{att},V}$ and $E_{\rm att}(B-V)$ enhances the total-to-selective attenuation $R_{\mathrm{att},V}$ compared to its extinction curve counterpart $R_{\mathrm{ext},V}$, with this discrepancy becoming particularly pronounced in massive edge-on galaxies under extreme optically thick conditions.

## 4 DISCUSSION

### 4.1 Intermediate-scale structures (ISS)

The extinction curve exhibits ISS such as the prominent 2175 Å UV bump. These features exhibit widths intermediate between narrow DIBs and large-scale extinction variations. Our analysis reveals tentative detections of intermediate-scale features at 4870, 6370, and 7690 Å in attenuation curves. These features align with recent detections in Milky Way extinction curve (D. Massa et al. 2020; G. M. Green et al. 2025 ; R. Zhang et al. 2024), but lack counterparts at the Galactic feature positions of 4370 and 5400 Å. To isolate broad-band attenuation properties from these localized substructures, we mask 150 Å windows centred on suspected ISS positions during curve fitting.

Three key arguments indicate these ISS features originate beyond terrestrial or Galactic foregrounds. First, their persistence at rest wavelengths contradicts expectations for atmospheric or Galactic foreground origins. Such origins would produce blueshifted features at varying positions due to redshift, ultimately cancelling out through stacking. Second, *Gaia* and *Hubble Space Telescope* spectral data used in Milky Way extinction studies (D. Massa et al. 2020; G. M. Green et al. 2025; R. Zhang et al. 2024) confirm the features originate from Galactic dust rather than terrestrial atmospheric absorption. Third, our SDSS extragalactic sample resides in high Galactic latitude regions with low extinction, where foreground dust insufficient to explain the observed ISS strength relative to the local extinction levels.

Although supplementary tests using EMILES stellar population templates (A. Vazdekis et al. 2016) reproduce ISS signals at the same rest wavelength (Appendix B), the definitive nature of these features requires further verification. We cannot fully exclude systematic biases arising from stellar population template selection or incompleteness in stellar spectral libraries. Future investigations requiring precise stellar population templates will be essential for verifying whether these ISSs are intrinsic to extragalactic attenuation curves. A quantitative assessment of their strengths relative to the 2175 Å bump and possible trends with stellar mass, inclination, or metallicity is deferred to future work.

### 4.2 Geometric effect

In Fig. 4, we observe that edge-on galaxies exhibit flatter attenuation curves. This phenomenon has been reported similarly in numerous observational studies (e.g. J. Chevallard et al. 2013; J. Leja et al. 2017; S. Salim et al. 2018). Recent high-resolution simulations and observational works aim to quantitatively disentangle the effects of dust-star geometry, scattering, and grain size distribution on the attenuation curve (e.g. Y. Dubois et al. 2024; K. Matsumoto et al. 2025; I. Shivaei et al. 2025).

Given the wavelength range and methodology of this work, we adopt a simplified toy model of a uniform dust-star mixture to provide a first-order quantitative interpretation of this geometric effect. Figs 5 and A1 in Appendix A show that the inferred extinction curve slope is invariant with inclination. This invariance isolates the first-order physics that maps extinction to attenuation: dust geometry. Our use of a uniform mixture is motivated by several aspects of the data and analysis. First, within our optical wavelength range, the flux is predominantly affected by the diffuse ISM dust acting on older stellar populations, with minimal contribution from clumpy star-forming regions (except in extreme edge-on cases) where significant differences in grain size distribution might exist between the diffuse and clumpy phases. Consequently, scattering in optical is largely isotropic thus acting similarly to absorption. In contrast, forward scattering in the UV band facilitates the escape of ultraviolet photons. Second, median stacking averages object-to-object differences, so the composite curve reflects mean dust and geometry. Most importantly, the inclination invariance of the extinction slope is both a result recovered from the attenuation curve fitting under the uniform mixture geometry and a core tenet of the model itself, which assumes intrinsic dust properties are isotropic. The consistency between our derived extinction curve slopes and this fundamental assumption provides strong support for the validity of the uniform mixture model as a first-order description.

The uniform mixture model captures two relevant first-order observational trends. Both the colour excess $E_{\rm att}(B-V)$ increases with inclination and the attenuation curve slope flattens toward inclination. This is fundamentally a saturation effect in a mixed geometry, where at high optical depths (edge-on view), the emitted flux is increasingly dominated by the outer, less obscured layers at high optical depth when galaxy is edge-on. When inverted to infer the intrinsic extinction law, the best-fitting slope remains nearly invariant with inclination up to $\theta < 75^\circ$, indicating that the inclination dependence of the attenuation curve is predominantly geometric rather than compositional.

We note deviations from this trend at high inclinations ($\theta > 75^\circ$) in massive discs. These likely mark the breakdown of a single-component uniform mixture under extreme optical depths and complex morphologies, and they motivate an explicit caveat: while the uniform mixture provides a useful first-order description, other effects can still play a role. Differential reddening between stellar populations of different ages is known to steepen/flatten attenuation slopes (e.g. A. K. Inoue 2005), and even within the Milky Way the extinction curve varies along different sight lines (e.g. E. L. Fitzpatrick & D. Massa 2007). Partial screen components from optically thick dust lanes, enhanced clumpiness and patchiness in dense ISM, and bulge-disc mixing within the fiber aperture can all modify the effective curve beyond a homogeneous mixture (J. W. Trayford et al. 2020; L. Sommovigo et al. 2025). Furthermore, dust scattering, which is not explicitly included in our simple model and tends to flatten the attenuation curve (K. Matsumoto et al. 2025; I. Shivaei et al. 2025), could also contribute. Restricting to $\theta < 75^\circ$ preserves the invariance of inclination of the inferred extinction slope across the stellar mass, supporting our main conclusion. Nevertheless, future work with multicomponent (birth cloud + diffuse ISM) or clumpy radiative transfer models with dust scatter effect (e.g. S. Charlot & S. M. Fall 2000; K.-I. Seon & B. T. Draine 2016; J. Lu et al. 2022), together with spatially resolved spectroscopy, will be necessary to quantify these higher order geometry effects.

### 4.3 Extinction slopes, grain size, and their evolution

In Section 3.2, we use a uniform dust-star mixture model to derive the intrinsic extinction curves from the observed attenuation. Our analysis reveals that, after removing geometric effects, the optical slope of the extinction curve systematically steepens with increasing stellar mass $M_*$ (equivalently, $R_V$ decreases with $M_*$), indicating that it reflects an evolution in the intrinsic dust properties rather than geometric projection.

Studies of attenuation curve slopes have reported conflicting results: some find steepening (e.g. A. J. Battisti, D. Calzetti & R. R. Chary 2017), others find no trend (e.g. M. Boquien et al. 2022; R. Fisher et al. 2025), and some find flattening (e.g. N. A. Reddy et al. 2018; S. Salim et al. 2018; I. Barišić et al. 2020) with increasing stellar mass. We suggest that these discrepancies arise because different studies use different wavelength ranges, tracers, and methods that are sensitive to distinct physical components. For example, the UV slope (e.g. I. Barišić et al. 2020) is primarily influenced by clumpy circumstellar dust around the youngest OB-type stars within dense birth clouds. The UV-optical slope (e.g. S. Salim et al. 2018) is more strongly affected by differential attenuation between young and old stars, which is often dominated by dust-star geometry (e.g. S. Charlot & S. M. Fall 2000; J. W. Trayford et al. 2020). In contrast, the optical slope of extinction curve derived in this work, based on the optical range (4200–8000 Å) and corrected for geometry, primarily constrains the properties of diffuse ISM dust affecting these stellar populations outside the clumpy regions. Therefore, the 'slope' measured in different studies actually probes different physical components, making direct comparisons inappropriate. Our results provide a systematic characterization of the mass-dependent evolution of intrinsic extinction in the diffuse ISM for a large sample of local galaxies.

The optical extinction slope is related to the dust grain size distribution, with a steeper slope (lower $R_V$) indicating a higher relative abundance of small grains, i.e. a higher small-to-large grain abundance ratio (SLR; e.g. B. T. Draine 2003; F. Galliano et al. 2018). Consequently, the mass-dependent steepening we report implies that the diffuse ISM of more massive galaxies is systematically enriched in small dust grains. This interpretation finds support in independent, multiwavelength studies. For instance, IR SED fitting reveals that the relative contribution of small dust grains increases under strong feedback conditions, leading to a higher fraction of small grains in the central regions of some massive galaxies with high SFR (M. Relaño et al. 2022; S. A. der Giessen et al. 2024).

Evidence from diverse astrophysical environments suggests a common regulator for this mass-dependent trend: the interstellar environment control over the dust grain size distribution. Observations of feedback-dominated regions, such as supernova remnants (R. S. Hill et al. 1995; R. Amanullah et al. 2014), the LMC2 supershell (K. D. Gordon et al. 2003), and the Milky Way bulge (D. M. Nataf et al. 2013), typically reveal low $R_V$ values, indicative of efficient grain fragmentation. In contrast, higher $R_V$ values signaling grain growth are predominantly found in denser, star-forming regions (G. De Marchi & N. Panagia 2014; Y. Wang, J. Gao & Y. Ren 2022a; Y. Wang et al. 2022b). This environmental dichotomy suggests that the balance between grain shattering and growth shifts toward fragmentation in environments characterized by strong interstellar radiation fields, shocks, and turbulence. The physical conditions in massive galaxies, with their deeper gravitational potentials, higher velocity dispersions, and cumulative feedback from past stellar activity, are highly conducive to this process. In such environments, frequent and sustained shocks can efficiently fragment large grains through catastrophic grain–grain collisions (M. Bocchio, A. P. Jones & J. D. Slavin 2014; A. P. Jones et al. 2017), while the high-kinetic energy and often lower density conditions of the diffuse ISM act to suppress the competing processes of grain coagulation and growth (R. S. Asano et al. 2013; R. McKinnon et al. 2021; N. H. Soliman et al. 2024). The systematic steepening of the extinction slope with stellar mass thus likely reflects a scenario where feedback-driven processing shifts the balance in the dust grain size distribution toward a higher proportion of small grains within the diffuse ISM.

However, this interpretation challenges some traditional expectations from dust evolution models. Simple models might predict that in the high-metallicity environments typical of massive galaxies, efficient grain growth and coagulation would lead to a dominance of large grains and flatter extinction curves (e.g. K.-C. Hou et al. 2019). Our finding of the opposite trend, steeper curves in more massive galaxies, suggests that within the diffuse ISM, the regulatory effects of galactic feedback (e.g. through shattering and the suppression of growth) can outweigh the influence of metallicity alone. This discrepancy underscores that the interplay between galactic ecosystems and dust evolution is more complex than often assumed. While our analysis cannot definitively rule out alternative explanations, the observed mass-dependent trend, clearly isolated from geometric effects, provides a key observational constraint. Future models of dust evolution must account for this apparent dominance of fragmentation processes within the diffuse ISM of massive galaxies, potentially by incorporating more realistic treatments of metal diffusion (L. E. C. Romano, K. Nagamine & H. Hirashita 2022) or more efficient fragmentation processes, such as ISM turbulence, feedback-driven shocks, and the distinct dust lifecycles in different galactic phases.

## 5 SUMMARY

In this paper, we presents a comprehensive investigation of the evolutionary dust attenuation curves in local SFGs using a non-parametric approach, the Stellar population synthesis with Equivalent Widths (SEW) method. By analysing 169 568 SFGs in SDSS DR7, we reconstruct wavelength-dependent attenuation curves free from parametric assumptions. Composite attenuation curves, constructed through median stacking across orthogonal stellar mass and inclination bins, reveal an evolutionary trend: higher stellar mass galaxies exhibit steeper attenuation slopes (lower $R_V$), while edge-on galaxies display flatter curves due to geometric saturation effects from increased dust column density.

Radiative transfer modelling under a uniform dust-star mixture confirms that the observed slope variations stem from intrinsic differences in dust properties rather than geometry effects alone. Critically, the extinction curve slopes derived from radiative transfer remain invariant with inclination angle at fixed stellar mass, demonstrating that the intrinsic dust grain properties are independent of viewing geometry. This inclination-insensitive behaviour of extinction slopes reinforces the interpretation that systematic $R_V$ variations with mass arise from dust evolution processes intrinsic to galaxies, rather than saturation effects. The steepening of extinction slopes with stellar mass aligns with mass-dependent dust processing mechanisms, where enhanced feedback in massive galaxies promotes grain shattering and suppresses coagulation, favouring smaller grain populations.

Tentative detections of ISS features at 4870, 6370, and 7690 Å challenge conventional attenuation prescriptions, suggesting unresolved complexities in dust composition or spatial distribution. These features persist across stellar population templates, though their definitive origin requires further validation.

The work bridges observed attenuation trends to underlying extinction properties, demonstrating that galactic-scale radiative transfer effects amplify differences in attenuation curve slope compared to pure extinction curves. The correlation between extinction curve slope and stellar mass mirrors environmental variations in local galaxies, linking dust evolution to galactic-scale processes. These findings indicate the dynamic interplay between dust geometry, grain size evolution, and galactic environment, providing constraints for models of dust lifecycle and ISM evolution.

## ACKNOWLEDGEMENTS

This work is partly supported by the National Key Research and Development Program (nos 2020SKA0110100, 2022YFA1602903, 2023YFB3002502), the National Natural Science Foundation of China (nos 12403016, 12533007), the Postdoctoral Fellowship Program of CPSF (no. GZC20241514), and the science research grants from the China Manned Space project with CMS-CSST-2025-A07, CMS-CSST-2025-A10. SS thanks research grants from the National Key Research and Development Program of China (no. 2022YFF0503402) and the National Natural Science Foundation of China (no. 12141302). SF acknowledges support from National Natural Science Foundation of China (no. 12103017), Natural Science Foundation of Hebei Province (A2025205037).

Funding for the SDSS and SDSS-II has been provided by the Alfred P. Sloan Foundation, the Participating Institutions, the National Science Foundation, the U.S. Department of Energy, the National Aeronautics and Space Administration, the Japanese Monbukagakusho, the Max Planck Society, and the Higher Education Funding Council for England. The SDSS Web Site is http://www.sdss.org/.

The SDSS is managed by the Astrophysical Research Consortium for the Participating Institutions. The Participating Institutions are the American Museum of Natural History, Astrophysical Institute Potsdam, University of Basel, University of Cambridge, Case Western Reserve University, University of Chicago, Drexel University, Fermilab, the Institute for Advanced Study, the Japan Participation Group, Johns Hopkins University, the Joint Institute for Nuclear Astrophysics, the Kavli Institute for Particle Astrophysics and Cosmology, the Korean Scientist Group, the Chinese Academy of Sciences (LAMOST), Los Alamos National Laboratory, the Max-Planck-Institute for Astronomy (MPIA), the Max-Planck-Institute for Astrophysics (MPA), New Mexico State University, Ohio State University, University of Pittsburgh, University of Portsmouth, Princeton University, the United States Naval Observatory, and the University of Washington.

## DATA AVAILABILITY

The data used in this paper are publicly available data from SDSS, MPA-JHU, and L. Simard et al. (2011). The SEW method is the publicly available pPXF-SEW.

## APPENDIX A: FROM CCM EXTINCTION CURVE TO ATTENUATION

This appendix employs the CCM extinction curve (J. A. Cardelli et al. 1989) within a uniform mixture geometry framework. The stellar mass dependence of CCM extinction curve slope $R_{55,\mathrm{CCM}}$ is presented in Fig. A1 with blue solid line:

$$R_{55,\mathrm{CCM}} = 0.38x^3 - 0.12x^2 - 0.97x + 3.47. \quad (A1)$$

The blue dashed line indicating the corresponding $R_{55}$-stellar mass relation for the power-law extinction curve originally shown in Fig. 5. The principal objective of this study focuses on investigating the systematic dependence of extinction curve slopes on stellar mass and inclination. Comparative analysis between these appendix results and the main text reveals consistent variation patterns across different curve formalisms: the extinction curve slope demonstrates negligible correlation with inclination angle but exhibits progressive steepening with increasing stellar mass. Although the CCM formalism consistently exhibits a weaker mass-dependent steepening trend compared to that of the power-law curve (indicating model curve dependent variations in absolute extinction measurements), such formalism-specific discrepancies lie beyond the scope of our current investigation. This consistency in mass-dependent behaviour across different curve implementations reinforces our primary conclusion regarding the fundamental role of stellar mass in shaping dust extinction properties.

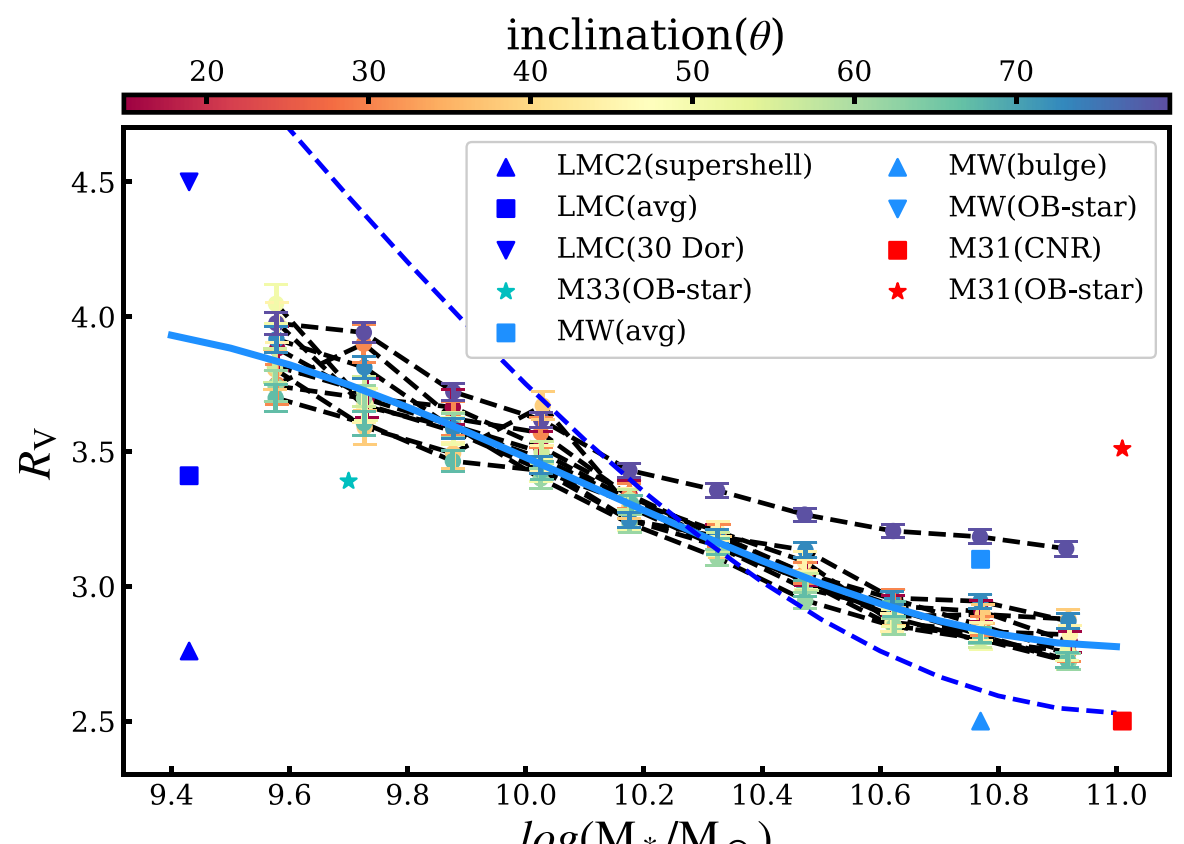

**Figure A1.** Same with the bottom panel of Fig. 5 but with fit CCM extinction curve, the light blue line represent the $R_{55}$-stellar mass relation of CCM extinction curve. While the blue dashed line is the $R_{55}$-stellar mass relation of power-law extinction curve in Fig. 5.

## APPENDIX B: ATTENUATION WITH EMILES SSPS

In this Appendix, the stellar population synthesis used in SEW-pPXF is the Emiles stellar population templates (A. Vazdekis et al. 2016) with a Chabrier IMF (age ranging from 30 Myr to 13 Gyr, metallicity ranging from −2.3 to 0.3). The stacked attenuation curve results of different stellar mass and inclination bins are shown in Fig. B1. The attenuation curves are similar to those derived from GALAXEV stellar population templates, with the ISSs observed at the same wavelength 4870, 6370, and 7690 Å.

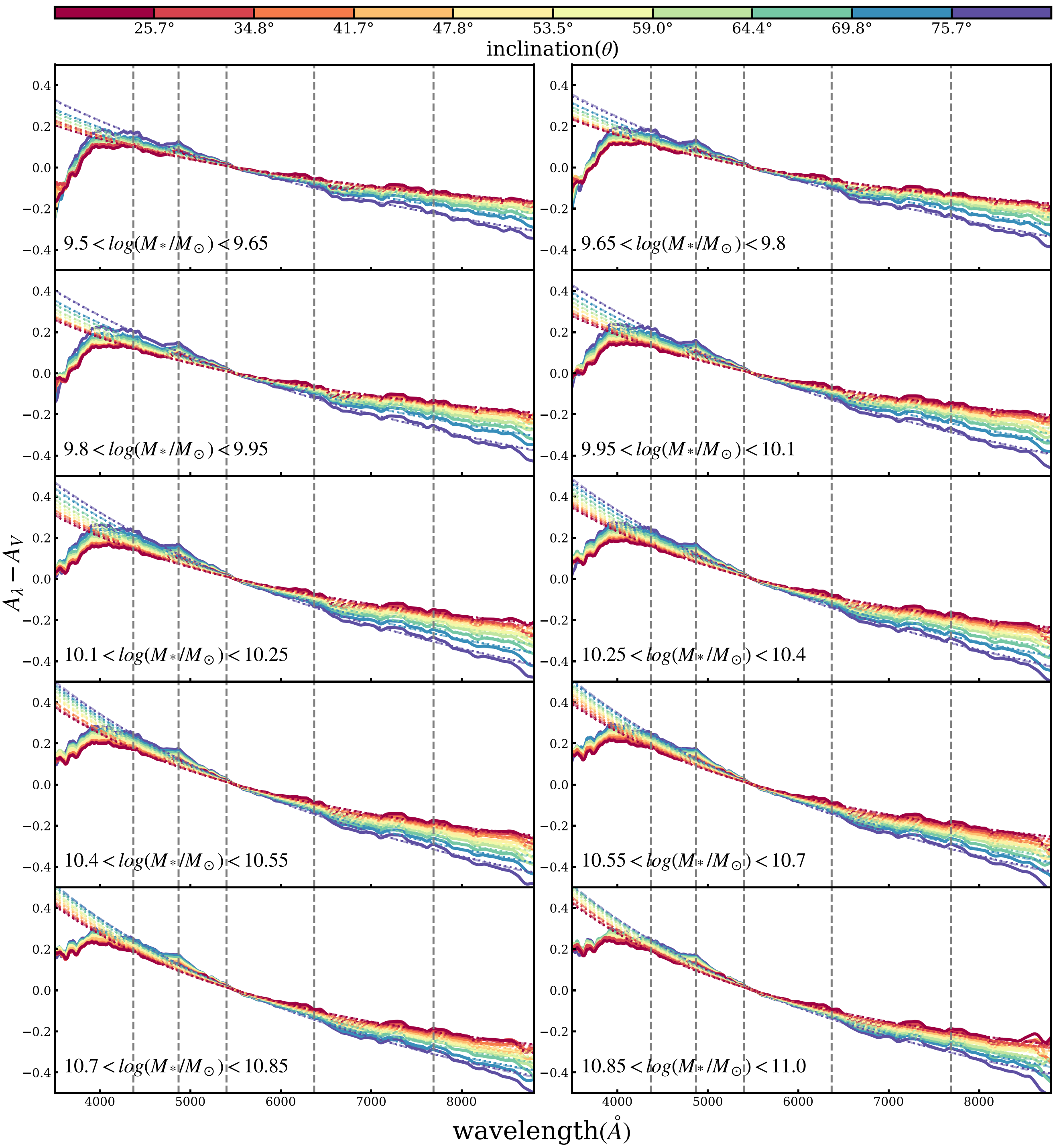

**Figure B1.** The same with Fig. 2 but with Emiles stellar population templates.